\documentclass[twocolumn]{aastex631}

\newcommand{\mhalo}{M$_{\rm halo}$}

\newcommand{\msun}{M$_\odot$}

\newcommand{\rs}{{\ifmmode{r_{\rm s}}\else{$r_{\rm s}$}\fi}}

\newcommand{\Eq}[1]{Eq.~(\ref{#1})}

\newcommand{\beq}{\begin{equation}}
\newcommand{\eeq}{\end{equation}}
\def\beqa{\begin{eqnarray}}
\def\eeqa{\end{eqnarray}}

\shorttitle{A shallow dark matter halo in a UDG}
\shortauthors{Brook et al.}
\graphicspath{{./}{figures/}}
\begin{document}

\title{A shallow dark matter halo in Ultra Diffuse Galaxy AGC 242019: \\ are UDGs  structurally similar to low surface brightness galaxies?}

\author{Chris B. Brook}
\affiliation{Universidad de La Laguna. Avda. Astrof\'{i}sico Fco. S\'{a}nchez, E-38200, La Laguna, Tenerife, Spain}
\affiliation{Instituto de Astrof\'{i}sica de Canarias, Calle Via L\'{a}ctea s/n, E-38206 La Laguna, Tenerife, Spain}

\author{Arianna Di Cintio}
\affiliation{Universidad de La Laguna. Avda. Astrof\'{i}sico Fco. S\'{a}nchez, E-38200, La Laguna, Tenerife, Spain}
\affiliation{Instituto de Astrof\'{i}sica de Canarias, Calle Via L\'{a}ctea s/n, E-38206 La Laguna, Tenerife, Spain}

\author{Andrea V. Macci\`{o}}
\affiliation{New York University Abu Dhabi, PO Box 129188, Saadiyat Island, Abu Dhabi, United Arab Emirates}
\affiliation{Center for Astro, Particle and Planetary Physics (CAP), New York University Abu Dhabi}
\affiliation{Max-Planck-Institut f\"{u}r Astronomie, Konigst\"{u}hl 17, 69117 Heidelberg, Germany}

\author{Marvin Blank}
\affiliation{New York University Abu Dhabi, PO Box 129188, Saadiyat Island, Abu Dhabi, United Arab Emirates}
\affiliation{Center for Astro, Particle and Planetary Physics (CAP), New York University Abu Dhabi}
\affiliation{Institut f\"{u}r Theoretische Physik und Astrophysik, Christian-Albrechts-Universit\"{a}t zu Kiel, Leibnizstr. 15, D-24118 Kiel, Germany}



\begin{abstract}
A central question regarding Ultra Diffuse Galaxies (UDGs) is whether they are a separate category  to Low Surface Brightness (LSB) galaxies, or just their natural continuation  towards low stellar masses.  In this letter, we show that  the rotation curve of the gas rich UDG AGC 242019 is well fit by a dark matter halo with inner slope that asymptotes to $\sim$-0.54, and that such fit provides a concentration parameter that matches theoretical expectations. This finding, together with previously works in which shallow inner profiles are derived for UDGs,  shows that the structural properties of these galaxies are like  other observed LSBs. UDGs  show slowly rising rotation curves and this favours formation scenarios in which internal processes, such as SNae driven gas outflows, are acting to modify UDGs profiles. 
\end{abstract}



\section{Introduction} \label{sec:intro}
Low surface brightness galaxies  (LSBs) generally show slowly rising rotation curves (e.g. \citealt{impey97,bothun97}). Thanks to their low baryonic content, it is possible to use the gas dynamics to faithful estimate their dark matter distribution, which turn out to be best fit by models with a  flat inner density (e.g. \citealt{deblok01,deblok08,oh11b,lelli16,katz16}). Empirically based models for such profiles have largely used inner ``cores" which asymptote to a slope of zero at the lowest radii (\citealt{burkert95}). By contrast, dark matter halos that form within gravity only cosmological simulations have ``cuspy" inner density profiles, which asymptote to a slope of minus one (\citealt{navarro96}, the ``NFW" profile).   

More recent theoretical models based on hydrodynamical simulations within cold dark matter cosmology have shown that outflows of gas can cause the expansion of dark matter (e.g. \citealt{Navarro1996, Read05, governato10, maccio12,Pontzen12}). Such models  allow a larger range of inner densities, which depend primarily on the stellar and total mass of the galaxy (\citealt{DiCintio2014a,DiCintio2014b,tollet15,chan15}). Such mass dependent density profiles have been shown to provide better fits to the rotation curves of observed galaxies than the self-similar  profiles found in N-body simulations (\citealt{katz16,lazar20}).

Ultra diffuse galaxies (UDGs)  are an interesting class of galaxies, not only due to the difficulties in finding (e.g. \citealt{bothun91,bothun97, vandokkum15,roman17}) and analysing them (e.g. \citealt{ruizlara18,forbes21}), but because their diffuse baryonic component  allows the underlying dark matter density profiles to be well constrained, particularly when there is sufficient gas within a disk that allows the measurement of rotation velocities (\citealt{pina20,shi21}). Various claims have been made about unusual properties of UDGs, such as unusually large (\citealt{vandokkum16})  or small (\citealt{vandokkum18}) quantities of dark matter  (see however \citealt{Trujillo19} and \citealt{Montes20} for a rebuttal of these claims), and the fact that some UDGs seem to fall off the baryonic Tully-Fisher relation (\citealt{pina20}).

Different formation mechanisms have been proposed for UDGs and more generally for LSBs. Scenarios that invoke environmental effects, such as tidal heating and ram pressure stripping, are most relevant for UDGs found within clusters or groups \citep{Fangzhou19,Martin19,Carleton19,Tremmel20,Sales20}, while for  isolated field UDGs it has been proposed that they may form in the high spin tail haloes of a regular LSB dwarf  population \citep{amorisco16}, or  that fluctuations in the gravitational potential, driven by galactic outflows,  could be able to expand the stellar populations within such galaxies (\citealt{dicintio17,chan18} see also \citealt{teyssier13}). Finally, some particular merger configurations can add angular momentum to the disk and/or a temporary boost in spin, and cause star formation to be redistributed to the outskirts of galaxies \citep{DiCintio19,Wright21}.

These formation mechanisms would presumably predict  different dark matter density profiles. If caused by high spin, the process of accreting high angular momentum gas would not affect the underlying dark matter. However, one may expect a low concentration halo in such cases \citep{maccio07} although the existence of a relation between spin and concentration is not entirely settled: it has been suggested the relation may be due to unrelaxed systems (\citealt{neto07,maccio07}).  Regarding the other scenario, if a stellar distribution had been affected by feedback processes, one could expect the dark matter to also be expanded. This means that it is interesting to measure the underlying mass profiles of UDGs in order to discriminate amongst formation scenarios.

Furthermore, given that it is well established that LSB galaxies tend to have relatively flat inner density profiles,  it is important to determine whether UDGs have similar properties, which would favour a close relation between these two classes of galaxies, and may indicate that they are better considered as a single class of galaxies (see \citealt{mcgaugh96}).

 A recent study by \citealt{shi21}, hereafter Shi21, has claimed that the gas rich UDG AGC 242019,  identified within the ALFALFA survey of HI galaxies \citep{Leisman17},  has a well resolved rotation curve that is best fit by a cuspy NFW profile. However, their recovered fit provides a concentration of the dark matter halo as low as 2, a value which is more than 5$\sigma$ off of the predicted c-M relation at such masses (e.g. \citealt{dutton14}).
 
In this letter, we show instead that galaxy AGC 242019 has a rotation curve well fitted by a dark matter halo with inner slope that asymptotes towards $\sim$-0.54, and that such fit provides a concentration parameter that is far closer to what is theoretically expected. This finding, together with previously reported works in which a shallow inner profile is derived for UDGs \citep{Leisman17,vanDokkum19} favours a formation scenario in which inner processes are acting to modify the UDGs inner profiles \citep{dicintio17,chan18}, and confirms that UDGs are  like  observed LSBs: they show slowly rising rotation curves.

The paper is organised as follows:  data and methods are introduced in Section 2, the results are presented in Section 3, while a  conclusions is discussed in  Section 4.

\section{Data and Methods} \label{sec:data}

 Found by the Arecibo Legacy Fast ALFA (ALFALFA) survey of HI galaxies (\citealt{Leisman17}), AGC 242019 is a UDG with stellar mass 1.37$\times$10$^8$M$_\odot$,  HI mass 8.51$\times$10$^8$M$_\odot$ and star formation rate  8.2$\times$10$^{-3}$M$_\odot$ yr$^{-1}$. 

The rotation velocity for the galaxy was derived by Shi21 from HI data using the Karl G. Jansky Very Large Array,  and H$\alpha$ data from the Wide-Field Spectrograph on the Australian National University 2.3 m telescope. Details of the data and modelling are found in Shi21, who applied a tilted-ring model to the HI 3-D datacube with 3DBarolo (\citealt{diteodoro15}) and combined this with a rotation curve derived from the H$\alpha$ data by assuming the same ring parameters. In this study we simply take the derived rotation curve and corresponding density profile at face value. We use the fiducial model of Shi21, but our results are qualitatively robust to to their models that use different mass to light ratios, distances and disc heights, with only small quantitive changes. 

We then use a double power law density profile  \citep{Jaffe83,Merritt06},  that has  been shown to provide excellent results on a large variety of simulated and observed galaxies \citep{DiCintio2014a,katz16}, to fit the data:

\beq
\rho(r)=\frac{\rho_s}{\left(\frac{r}{\rs}\right)^{\gamma}\left[1 +\left(\frac{r}{\rs}\right)^{\alpha}\right]^{(\beta-\gamma)/\alpha}}
\label{equation}
\eeq
 
 \noindent where $\rs$ is the scale radius and $\rho_s$ the scale density.
  $\rs$ and $\rho_s$ are characteristics of each halo, related to their mass and formation time
\citep[e.g.][]{bullock01,maccio07}.  The inner and
outer regions have logarithmic slopes $-\gamma$ and $-\beta$,
respectively, while $\alpha$ regulates how sharp the transition
  is from the inner to the outer region. The NFW profile
  has $(\alpha,\beta,\gamma)=(1,3,1)$.

\begin{figure}
\includegraphics[width=3.3in,height=2.2in]{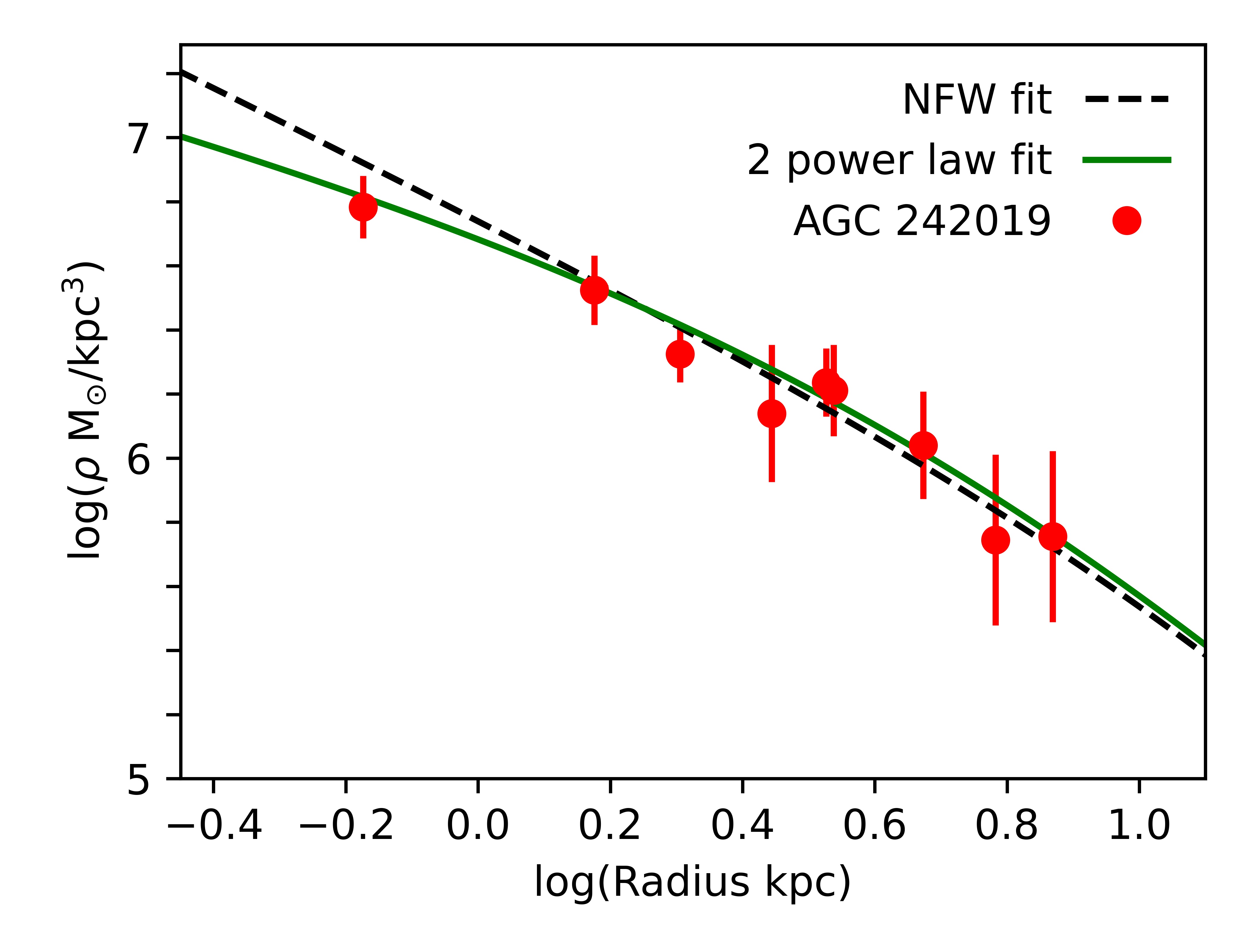}
\caption{The density profile of dark matter for AGC 242019, shown as red points. The green line is the double power law fit with central slope $\gamma$=-0.54, while the dashed black line shows the low concentration NFW fit derived in \citet{shi21}.
}
\label{fig:density} 
\end{figure}

Results are then compared to simulation data from the MaGICC (\citealt{brook12b}), NIHAO (\citealt{wang15}) and FIRE-2 (\citealt{hopkins18}) zoom in hydrodynamical cosmological galaxy formation simulations.

%%%%%%%%%%%%%%%%%%%%%%%%%%%%
\section{Results} \label{sec:results}
%%%%%%%%%%%%%%%%%%%%%%%%%%%%

 Figure~\ref{fig:density} shows the DM density profile for AGC 242019 taken from Shi21, as red circles. Our two power law fit is shown as a solid  green line. The fit has M$_{\rm{halo}}$=2.88$\times$10$^{10}$\msun, scale radius \rs = 6.4 kpc, concentration $c$ = 10.1,  inner slope $\gamma$ = 0.54,  outer slope $\beta$ = 2.15 and  transition parameter $\alpha$ = 0.89. The NFW fit derived in Shi21 is shown as a dashed black line, which has M$_{\rm{halo}}$ = 3.5$\times$10$^{10}$\msun,   \rs = 33.3 kpc and  $c$ = 2.0. The reduced $\chi^2$ for the power law and NFW fits were 0.2 and 0.5 respectively.  

\begin{figure}
\includegraphics[width=3.3in,height=2.2in]{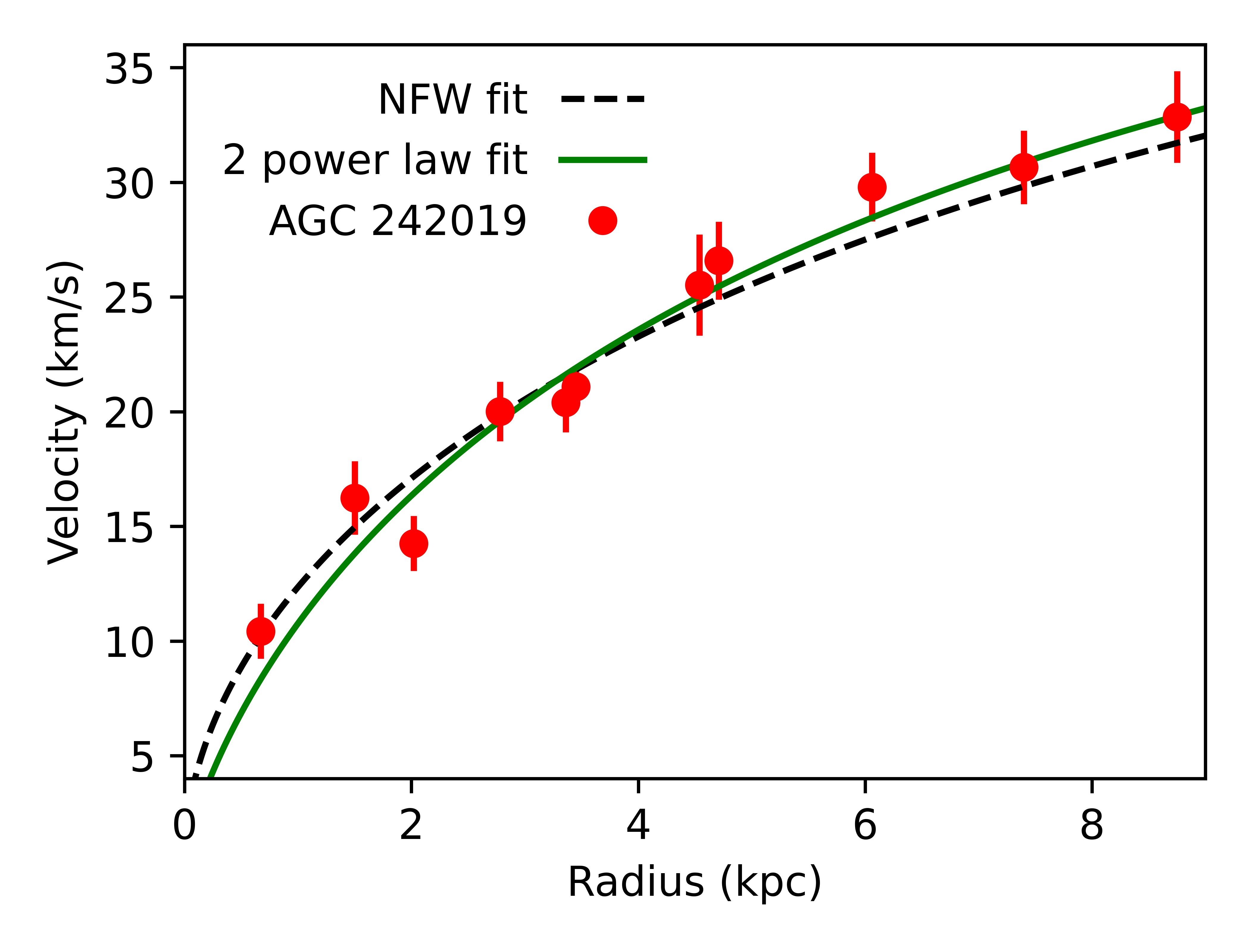}
\caption{The rotation curve of the dark matter for AGC 242019, shown as red points. The green line is the double power law density profile with central slope $\gamma$=-0.54, while the dashed black line shows the  low concentration  NFW fit derived in \citet{shi21}.
}
\label{fig:rc} 
\end{figure}

Figure~\ref{fig:rc} shows the dark matter contribution to the rotation curve for AGC 242019, also taken from Shi21, with the two power law fit from above shown as a solid  green line and the NFW fit from Shi21, as the dashed black line. The reduced $\chi^2$ for the two power law and NFW fits were 1.5 and 1.6 respectively.  

Of course, the two power law fit has more free parameters than the NFW profile, and so one can expect a better fit.  Further, with five  free parameters, the two  power law fit is known to be  degenerate (\citealt{klypin01}).  We are not claiming here that our fit is unique. Indeed, by fixing $\alpha$ =1 and leaving $\beta$  and $\gamma$ free, another set of parameters also made an approximately equally good fit, with
M$_{\rm{halo}}$ of 2.5$\times$10$^{10}$\msun, concentration $c$ = 8.3,  $\beta$=2.32,  and an asymptotic inner slope $\gamma$=0.55. If one uses a double power law profile with $\alpha$ and $\beta$ fixed to be 1 and 3 respectively (a profile sometimes as referred to as a generalised NFW profile) one can get another approximately equally good fit as for our fiducial two power law one, by using parameters M$_{\rm{halo}}$ of 2.0$\times$10$^{10}$\msun, concentration $c$ = 4.6 and asymptotic inner slope $\gamma$=0.57. A summary of the various  best fit parameters is provided in Table 1.

One can clearly see a trend in these fitting parameters. As we reduce the number of free parameters and fix the outer and then inner slope to match halos that form in dark matter only simulations, the way the parameters compensate to arrive to an acceptable fit is to decrease the value of the concentration. Once parameters are fixed at their  NFW values, i.e. $(\alpha,\beta,\gamma)=(1,3,1)$, then the concentration needs to be very low, c=2, in order for the  density profile and rotation curves to be reasonably fitted.

\begin{figure}
\includegraphics[width=3.2in,height=2.2in]{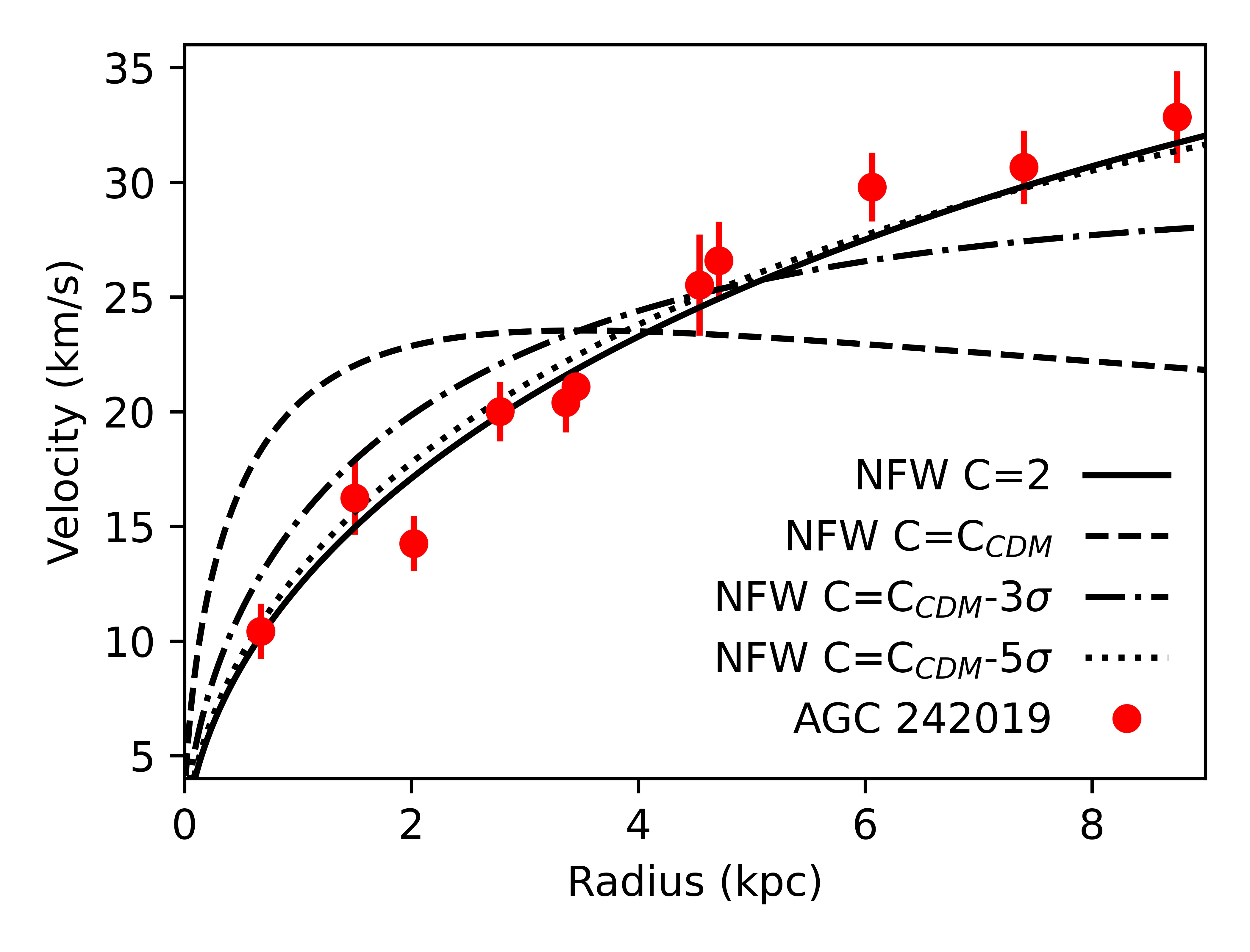}
\caption{The rotation curve for an NFW fit of various concentrations. The solid line shows the  NFW fit derived in \citet{shi21}. The dashed line shows the NFW fit when using the concentration expected from dark matter only simulations, C$_{\rm{CDM}}$, i,.e. following the mass-concentration relation of \citet{dutton14}. The dot dashed line shows the NFW fit using a concentration 3$\sigma$ lower than  C$_{\rm{CDM}}$, whilst the dotted line shows the NFW fit using a concentration 5$\sigma$ lower than  C$_{\rm{CDM}}$.}
\label{fig:concentration} 
\end{figure}

      The dependence of profile shape on concentration is demonstrated in Figure~\ref{fig:concentration}, that again shows the dark matter contribution to the rotation curve of  AGC 242019 as red circles, and then shows various NFW fits, firstly (dashed line) by fixing the concentration according to the mass-concentration relation from cosmological simulations (\citealt{dutton14}), then by fixing the concentration to be 3$\sigma$ lower than the expected relation (dot-dashed curve) and finally by being  5$\sigma$ below the expected relation (dotted curve).

\begin{table}
    \centering
    \begin{tabular}{lcccc}
    \hline
      Fit type   &   M$_{halo}$/\msun  &  c  & $\gamma$ & $\chi^2$\\
         \hline\hline
     
                 $(\alpha,\beta,\gamma)=(0.89,2.15,0.54)$ & 2.9$\times$10$^{10}$  & 10.1 & 0.54 & 0.2 \\
                                  $(\alpha,\beta,\gamma)=(1,2.32,0.55)$ & 2.5$\times$10$^{10}$  & 8.3 & 0.55 & 0.2 \\
                                   $(\alpha,\beta,\gamma)=(1,3,0.57)$ & 2.0$\times$10$^{10}$  & 4.6 & 0.57 & 0.2 \\
        $(\alpha,\beta,\gamma)=(1,3,1)$ & 3.5$\times$10$^{10}$  & 2.0 & 1 & 0.5 \\
                \hline
    \end{tabular}
    \caption{Fit parameters of the density profile of UDG AGC 242019, based on the general double power law in \Eq{equation}. From top to bottom are fits with progressively fewer  numbers of free parameters: in  the first row $(\alpha,\beta,\gamma)$ parameters were all free to vary, then $\alpha$ was fixed to a  value of 1, then $\alpha$  and $\beta$ were fixed  to be 1 and 3, respectively, and finally  $(\alpha,\beta,\gamma)$ were set to be (1,3,1), i.e. an NFW profile. In each case the recovered  halo mass, concentration, asymptotic inner slope and reduced $\chi^2$ are shown.}
    \label{tab:fit}
\end{table}

These  fits have  concentrations of 15.7, 5.7 and 2.7 and halo masses of  1.8e10$\times$10$^{9}$, 6.4$\times$10$^{9}$, 2.3$\times$10$^{10}$ \msun, respectively, for  concentration set to be    0$\sigma$, 3$\sigma$ and 5$\sigma$ away from the expected relation. The NFW fit to the data of Shi21 has a concentration, $c=$2.0, that is more than  5$\sigma$ below the expected concentration of halos of such mass, according to N-body only simulations. 
Lowering the concentration also moves the fitted halo to have a higher mass, which brings it into better agreement with abundance matching relations (e.g. \citealt{Moster13}).

Summarizing, reproducing at the same time reasonable fits to rotation curves, along with theoretically expected values of concentration and halo mass, is a challenge for NFW models when applied to UDGs and LSBs, which show slowly rising rotation curves.
Much better fits and c-\mhalo values can be simultaneously obtained when allowing for a shallow inner profile (see \citealt{katz16} for a full discussion of this issue). 

\begin{figure}
\includegraphics[width=3.2in,height=2.2in]{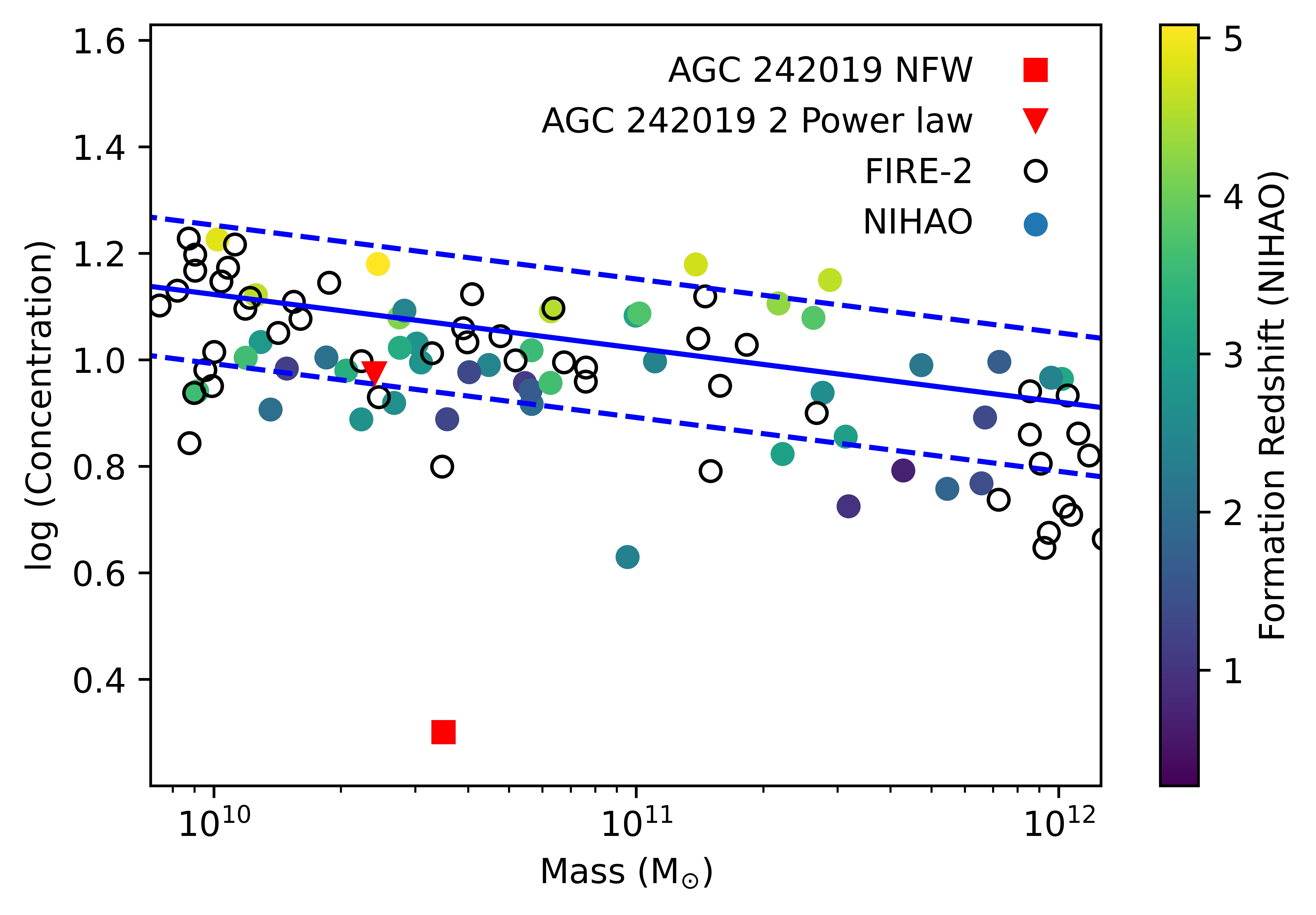}
\caption{The halo mass-concentration relation  and its  1$\sigma$ error are shown as blue solid and dashed lines, as derived in  \citet{dutton14} from dark matter only simulations. Hydrodynamical simulations: NIHAO galaxies are shown as solid circles, colored by their half-mass formation redshift, while FIRE-2 as open circles. The red square and triangle respectively show AGC 242019 for the  NFW fit derived in Shi21 and for the double power law fit used in this work, with central slope $\gamma$=0.54.}
\label{fig:Mconcentration} 
\end{figure}

In Figure~\ref{fig:Mconcentration} we explicitly  plot the  c-M$_{\rm{halo}}$ relation, with our derived values for UDG AGC 242019   using the two power law fit (red triangle) and for the NFW fit (red square). The relation  from N-body only simulations is shown as a solid blue line, with the 1$\sigma$ deviations shown as a blue dashed line (\citealt{dutton14}). FIRE-2 simulations  are shown as open  circles (\citealt{lazar20}). NIHAO simulations are shown as circles colored by  formation redshift. The trend for later forming galaxies to have lower concentrations, first  shown in \citealt{wechsler02}, is clear from the colors. 

%Shown as open circles are the N-body only simulations from NIHAO and FIRE-2 that correspond to the halos plotted from the fully hydrodynamical galaxy formation simulations. 

%Note that NIHAO and FIRE initial conditions are chosen to span the range of halo concentrations (and spins), rather than sample their distributions, so one can expect greater scatter than what is found in the fully sampled N-body simulations. 
 
 As expected from Figure~\ref{fig:concentration}, fitting AGC 242019 with an  NFW profile results in a concentration that is a long way from what is expected from N-body simulations, as well as being a long way from what is found in hydrodynamical simulations that model galaxy formation within a cosmological context. A shallow inner density profile provides concentration values in line with expectations.

\begin{figure}
\includegraphics[width=3.4in,height=2.2in]{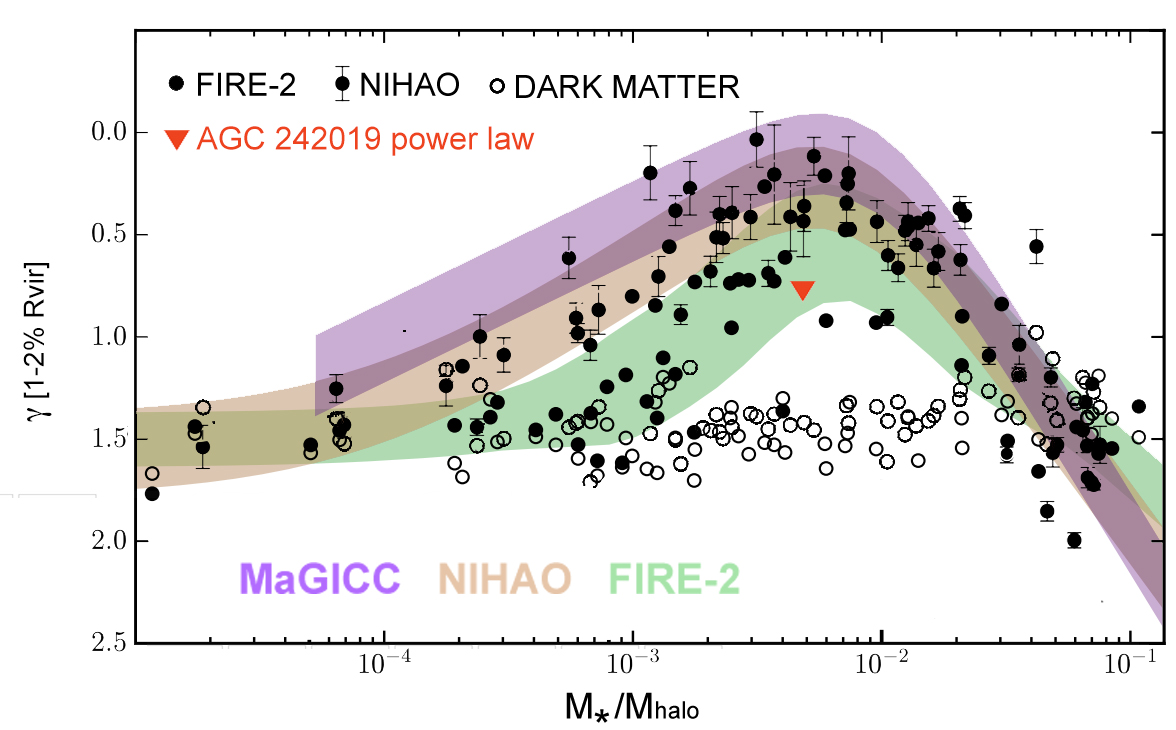}
\caption{The relation between the inner slope $\gamma$ of the dark matter density profile, measured between 1$\%$ and 2$\%$ of the virial radius ($\gamma_{12}$),  and the ratio of stellar to halo mass, M$_{\rm{star}}$/M$_{\rm{halo}}$. Fits from MaGICC (\citealt{DiCintio2014a}), NIHAO (\citealt{tollet15}) and FIRE-2 (\citealt{chan15}) simulations are shown as shaded regions with a scatter of $\Delta\gamma$=$\pm$0.2. Black dots with error bars are hydro NIHAO simulations,  black dots without error bars are from hydro FIRE-2, whilst black circles are dark matter only simulations from both NIHAO and FIRE-2. AGC 242019 is shown as an upside down triangle, as derived from the double power law fit used in this study, having $\gamma_{12}=0.78$ (asymptotic $\gamma=0.54$).}
\label{fig:abun} 
\end{figure}

Figure~\ref{fig:abun} shows the relation between the inner slope $\gamma$ of the density profile, measured between 1$\%$ and 2$\%$ of each galaxy virial radius ($\gamma_{12}$),  and the ratio of stellar to halo mass, M$_{\rm{star}}$/M$_{\rm{halo}}$. 
Resuls from MaGICC (\citealt{DiCintio2014a}), NIHAO (\citealt{tollet15}) and FIRE-2 (\citealt{lazar20}) simulations are shown as shaded regions, covering a range of $\Delta\gamma$=$\pm$0.2 from each average relation. Black dots with error bars are NIHAO hydro-simulations,  black dots without error bars are from FIRE-2, whilst open black circles are dark matter only simulations from NIHAO and FIRE-2. AGC 242019 is shown as an upside down triangle, derived from the double power law fit used in this study. The slope between 1$\%$ and 2$\%$ of the virial radius is larger than the asymptotic inner slope: in this case,  an asymptotic inner slope of  $\gamma$=0.54 translates into a   $\gamma_{12}$=0.78, which is represented in Figure~\ref{fig:abun}.
By comparison, N-body only simulations, which asymptote to $\gamma$= 1,  can be seen in the open circles to have values of $\gamma_{12}$$\approx$1.5. Our derived inner slope for AGC 242019  is around half as steep as expected for N-body simulations, regardless of the particular radial region adopted to compute $\gamma$, and in line with hydro simulations that predict an expanded profile at the mass range of this UDG.

%%%%%%%%%%%%%%%%%%%%%%%%%%%
\vspace{-.0cm}\section{Discussion and Conclusions} \label{sec:discussion}
%%%%%%%%%%%%%%%%%%%%%%%%%%%%
 A central question regarding UDGs is whether they are a separate category of galaxy than LSBs. Determining their properties, including their mass profiles, is therefore an important aspect to address. 

AGC 242019 is a field UDG,  with a large HI gas content that has allowed detailed rotation curve to be created and mass modelling to be carried out. \citet{shi21} found that the density profile of this galaxy is best fit by an NFW profile with M$_{\rm{halo}}$=3.5$\times$10$^{10}$\msun\ and a concentration $c$ = 2.0, the latter being more than 5$\sigma$ away from theoretical expectations.  In this letter we have instead shown that a better fit can be obtained employing  a two-power law model, which allows the inner slope of the dark matter halo to be shallower than a  cuspy profile. 

In our fit, we found  $(\alpha,\beta,\gamma)=(0.89,2.2,0.54)$ (where $\gamma$ and $\beta$ are the inner and outer slopes, and $\alpha$ regulates  the transition between them), a halo mass of  M$_{\rm{halo}}$=2.88$\times$10$^{10}$\msun\ and a concentration $c$ = 10.1, which is in line with expectations from c-M relations \citep{dutton14}.
We do not claim that our fit is unique. Indeed we suggested that by restricting the transition parameter to $\alpha$=1, or  by restricting   both $\alpha$=1 and the outer slope $\beta$=3, we can find almost equally good fits, in terms of  their reduced $\chi^2$. However, as more restrictions towards an NFW profile are included, resulting concentrations are forced to become progressively lower, up to the point of the extremely low c  (and higher $\chi^2$) reached in  Shi21, which does not match c-M predictions.

Are UDGs structurally similar to LSB galaxies? 
Like LSB galaxies, AGC 242019 has a slowly rising rotation curve, compared to that expected in N-body only simulations. One explanation is to fit the slowly rising rotation curve of AGC 242019 with a NFW profile with an extremely (more than 5$\sigma$ outlier) low concentration (Shi21).  In fact, \cite{neto07} found that the low concentration tail in N-body simulations is caused by unrelaxed systems. If such unrelaxed systems are excluded, the scatter in concentrations at a given mass becomes considerably smaller. Given that there is no evidence that  AGC 242019 has undergone a recent merger,  such a low concentration halo would be a considerably larger outlier from what is expected from N-body simulations, so far more than 5$\sigma$.  

Although the possibility of an extremely low concentration halo as proposed by Shi21 cannot be totally ruled out, it should be recalled that many/most LSB galaxies have slowly rising rotation curves: they cannot all be explained as being  extreme outliers from the c-M  relation. The alternative is that the dark matter halo in these galaxies has a less steep inner density profile and a concentration in line with theoretical expectation  (e.g. \citealt{DiCintio19} and references therein). When placed in this context, AGC 242019 is very typical of the well studied LSB population (e.g. \citealt{deblok08}), and can be considered as just a natural extension of LSBs at low stellar mass.

Regarding formation theories  for UDGs, a very late forming halo with extremely low concentration may fit with the theory of UDGs forming in high spin halos  (\citealt{amorisco16}), while interpreting the galaxy as being a normal LSB with a relatively shallow inner density profile, would fit will with the theory of expanded stellar and dark matter populations (\citealt{dicintio17,chan18}). We note, however, that the scenarios are not mutually exclusive. 

Other suggested formation scenarios that could be at play in both the field and in groups/clusters have been advocated to be able to explain the number density and global properties of UDGs, such as their color, effective radii, large metallicities and number densities \citep{Fangzhou19,Carleton19,Tremmel20,Sales20,Wright21}.

%Amongst these theories, we find the tidal heating scenario, in which group UDGs form by tidal ‘puffing up’  at pericenter passage and become later on quiescent by ram-pressure stripping of gas \citep[see also \citealt{Sales20} for implication of tidal heating on the UDGs metallicity]{Fangzhou19,Tremmel20},  the  merger-driven scenario, in which early mergers produce a temporary boost in spin, and cause star formation to be redistributed to the outskirts of galaxies, resulting in lower central star formation rates \citep{Wright21}, and  a combination of several of these effects, such as early star formation plus heating and stripping in clusters \citep{Martin19}, which is more effective if the galaxy had an initial dark matter central core  \citep{Carleton19}.
Formation mechanisms that are able to reproduce UDG global properties without the need of making a central shallow core,  fail at  reproducing the radial properties of UDGs' rotation curve. Evidence is beginning to suggest that UDGs have slowly rising rotation curve, just like the more general category of LSBs. For example,  UDGs AGC 122966, AGCs 219533 and AGC 334315, HI-rich objects from the ALFALFA survey, all show rotation curves and mass profiles compatible with having shallow central DM profiles \citep{Leisman17}; UDG DF44 shows a stellar velocity dispersion which is well fit by a mass-dependent cored profile from \citealt{DiCintio2014b}, which provides a mass of  M$_{\rm{halo}}$$\sim$$10^{11.2}$\msun, isotropic orbits and a positive kurtosis, in qualitative agreement with measurements \citep{vanDokkum19}. Furthermore, WLM is a Local Group galaxy that may be considered as a UDG: with a stellar mass of 1.1$\times$10$^7$\msun, and a  half light radius of  1.65 kpc, WLM has been found to have a slowly rising rotation curve and similarly shallow central dark matter profile  as AGC 242019 (\citealt{leung21}).

%\ARY{if u want to say that sims might not be good enough yet to fully reproduce the whole variety of uDG, just say it }It is important to recognise that if profiles can be affected by baryons, then a range of factors during the formation and evolution of galaxies could have an effect on the final profile, such as formation times, merger histories, baryon fractions, gas and stellar distributions, rates and distributions of star formation, and angular momentum distributions. These factors have not yet been fully explored within cosmological simulations , and indeed it is not clear that current simulations are sophisticated enough, and well enough resolved within a large enough statistical sample to fully explore these possible effects (although see \citealt{Read16} for an exploration of the importance of star formation history). If the halo is indeed expanded,  UDGs could still constitute the the high spin halo tail compare to  LSBs. In other words, the high spin and  stellar expansion scenarios may not be entirely mutually exclusive.

In conclusion, AGC 242019 has a similar mass distribution as other well studied LSB galaxies. 
Determining  the radial properties, in particular the  mass profiles, of a larger sample of UDGs is an important aspect to address, in order to better understand UDGs and their formation, as well as their connection to LSBs. If the structural properties of UDGs and LSBs are alike as it appears, then explaining their mass distribution  would require halo expansion by baryonic processes (e.g. \citealt{governato10,dicintio17,chan18}), or more exotic forms of dark matter (e.g. \citealt{Schive14,Zavala19}) or  gravity (e.g. \citealt{lelli17}).

\begin{acknowledgments}
CB  is supported by  grant  PGC2018-094975-C22 from the Spanish Ministry of Science and Innovation.
ADC is supported by a Junior Leader fellowship from `La Caixa' Foundation (ID 100010434), fellowship  LCF/BQ/PR20/11770010. Part of the research was carried out on the HPC resources at New York University Abu Dhabi. 
\end{acknowledgments}

\bibliographystyle{aasjournal}
\bibliography{archive}

\end{document}